\documentclass[preprint,showpacs,preprintnumbers,amsmath,amssymb]{revtex4}

\usepackage{graphicx}
\usepackage{dcolumn}
\usepackage{bm}

\begin{document}
\title{The roughness of stylolites:\\ Implications of 3D high
resolution topography measurements}

\author{J. Schmittbuhl}

\affiliation{Laboratoire de G{\'e}ologie, UMR CNRS 8538, Ecole Normale
Sup{\'e}rieure,\\ 24, rue Lhomond, F--75231 Paris C{\'e}dex 05,
France. Email: Jean.Schmittbuhl@ens.fr.}

\author{F. Renard$^\ddag$ and J.P. Gratier}

\affiliation{LGIT-CNRS-Observatoire, Universit\'{e} J. Fourier BP 53,
F-38041 Grenoble, France \& $^\ddag$Physics of Geological Processes, University
of Oslo, Norway.}

\author{R. Toussaint}

\affiliation{Institute of Physics, University of Oslo, PB 1048,
Blindern, N-0316 Oslo, Norway.}

\date{\today}
\begin{abstract} 
Stylolites are natural pressure-dissolution surfaces in sedimentary
rocks. We present 3D high resolution measurements at laboratory scales
of their complex roughness. The topography is shown to be described by
a self-affine scaling invariance. At large scales, the Hurst exponent
is $\zeta_1 \approx 0.5$ and very different from that at small scales
where $\zeta_2 \approx 1.2$. A cross-over length scale at around $\L_c
=1$~mm is well characterized.  Measurements are consistent with a
Langevin equation that describes the growth of a stylolitic interface
as a competition between stabilizing long range elastic interactions
at large scales or local surface tension effects at small scales and a
destabilizing quenched material disorder.
\end{abstract} 
\maketitle
Stylolites are geological patterns that are very common in polished
limestones, a material largely used to construct floors and walls of
buildings and monuments. They are observed as thin irregular
interfaces that look like printed lines on rock cuts, which is
responsible for their name. They are roughly planar structures that
are typically perpendicular to the geological load (i.e. lithostatic
pressure or tectonic maximum compressive stress). These rock-rock
interfaces are formed at shallow depths in the Earth's crust during
deformation of sedimentary rocks and result from a combination of
stress-induced dissolution and precipitation processes
\cite{Park68}. They are found in many sedimentary rocks such as
limestones, sandstones or evaporites \cite{Bathurst71} and exist on a
very large range of scales, from micro-meters to meters.

Despite their abundance, stylolites are, as mentioned by {\it Gal et
al.} \cite{Gal98}, {\it ``among the least well-explained of all
pressure-solution phenomena''}. First they are complex 3D structures
that are often only described from 2D cross-sections since they are
generally partially sealed \cite{Karcz03}. Second, they develop in
various geological contexts which lead to very different
geometries. Third they are sometimes transformed because of processes
like diagenesis and metamorphism that develop after their initiation.

In this Letter we show the first 3D high resolution topography
measurements of natural stylolite interfaces that could be fully
opened. We characterized the scaling invariance, namely self-affinity,
of the morphology and show the presence of a cross-over length
scale. We also propose a model of the stylolites roughening. It is
based on a Langevin equation that accounts for stress-induced
dissolution in a quenched disorder.

\begin{figure}[ht]
\begin{center}
\vspace*{5mm}\includegraphics[width=12cm,angle=0]{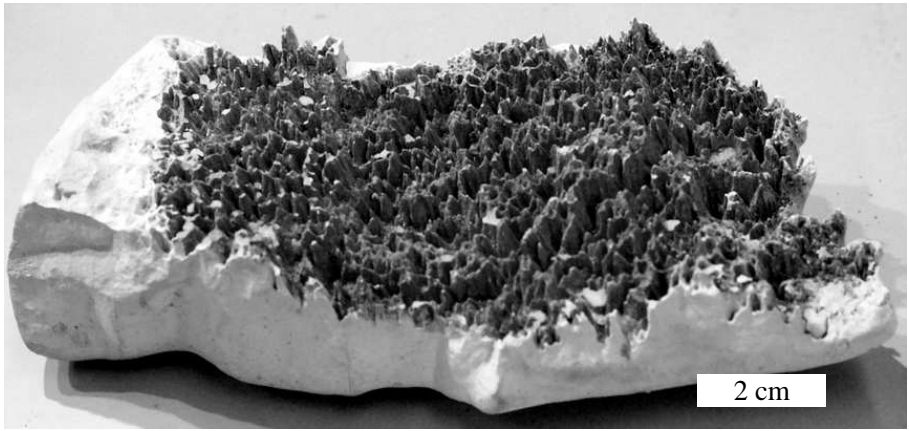}
\caption{\label{fig_stylo_sample} Picture of a stylolite surface
(S12A) in a limestone from Vercors Mountains. Magnitude of the peaks
are typically of the order of 6 mm.}
\end{center}
\end{figure}

The roughness measurements have been performed on three independent
stylolite interfaces included in very fine-grained limestone samples
from Burgundy area, Vercors, and Jura mountains in France
(Fig.~\ref{fig_stylo_sample}). The samples have been collected in
newly open quarries, thus preserved from late breakage and chemical
erosion. The opening procedure was possible for these samples because
of the accumulation of undissolved minerals like clays that formed a
weak layer along the stylolite interface. The concentration of these
minerals provides an estimate of the cumulative strain by dissolution
the sample underwent \cite{Renard03}. As shown in
Fig.~\ref{fig_stylo_sample}, peaks along the interface are randomly
distributed in space and of various sizes (up to one
centimeter). Large peak magnitudes and local high slopes along the
topography makes the roughness measurement difficult and challenging.

\begin{figure}[ht]
\begin{center}
\includegraphics[width=14cm,angle=0]{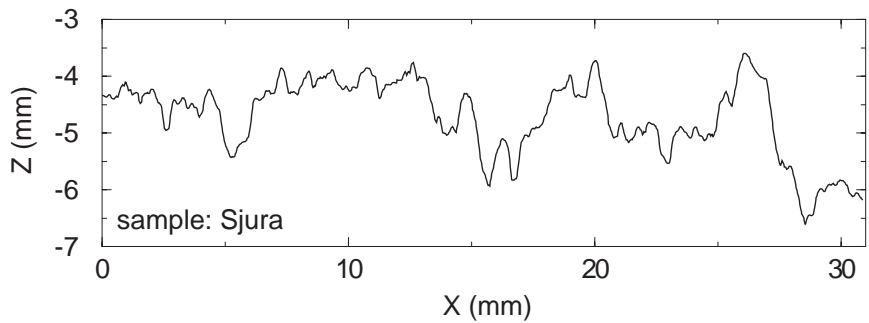}
\caption{\label{fig_roughness_prof} A 1D profile obtained by a
mechanical profilometer (1030 data points - $\Delta x=30\mu$m) along a
stylolite surface. }
\end{center}
\end{figure}
We used two different profilometers to sample the stylolite
roughness. First, with a mechanical profilometer
\cite{Schmittbuhl95,Lopez98} we extracted four profiles of 1030 points
each with a horizontal step of $\Delta x=30\ \mu$m. The mechanical
profilometer measures the surface height from the contact of a needle
onto the surface. The radius of the needle tip is 25 $\mu$m. The
vertical resolution is $3\ \mu$m over the available range of 5
$cm$. One of these profiles is shown in
Fig.~\ref{fig_roughness_prof}. We compare the mechanical measurements
to an optical profiling \cite{Meheust02}. This technique is based on a
laser triangulation of the surface without any contact with the
surface. The laser beam is $30\ \mu$m wide. Horizontal steps between
measurement points were $\Delta x=\Delta y = 7$ to $50\ \mu m$ with a
vertical resolution of $2\ \mu$m. The main advantage of this technique
comes from the high acquisition speed that can be performed compared
to the mechanical profilometer, since there is no vertical move and
on-flight measurements are possible. However, a successful comparison
with mechanical measurements is necessary to ensure that optical
fluctuations are height fluctuations and not material property
fluctuations. Three independent samples have been measured at very
high resolution: one side of a stylolite from Jura mountains (Sjura)
with a resolution $600 \times 600$, one side of a stylolite from
Burgundy area (S15) with a resolution $8200\times 4100$ and two
opposite surfaces of the same stylolite from Vercors mountains shown
in Fig.~\ref{fig_stylo_sample} with a resolution $2400\times 1400$ for
S12A and $8200\times 4100$ for S12B.

\begin{figure}[ht]
\begin{center}
\includegraphics[width=12cm,angle=0]{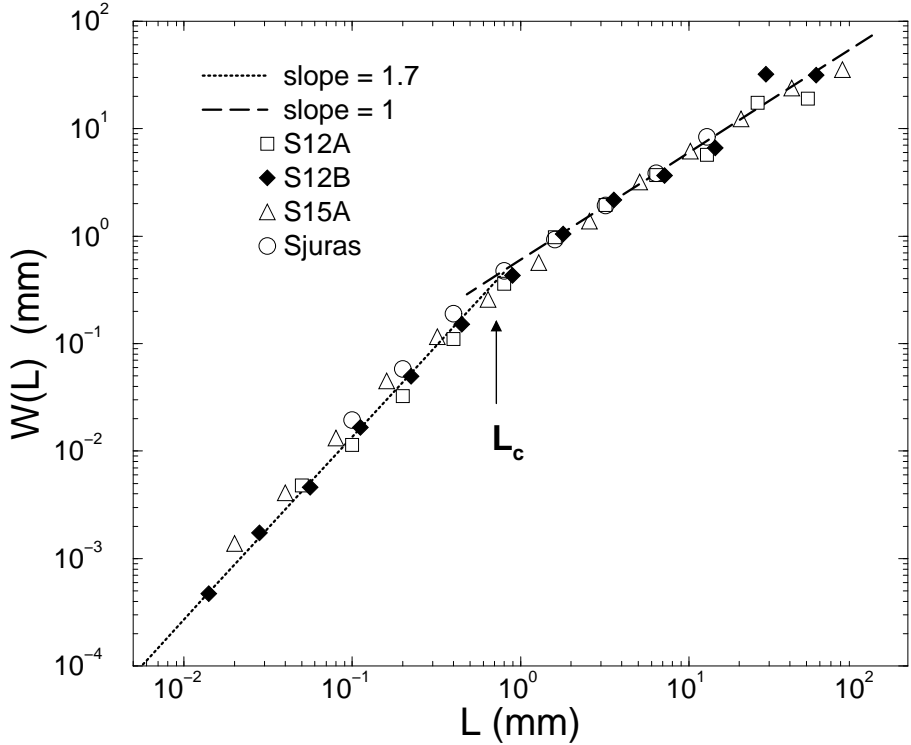}
\caption{\label{fig_spum_data} Averaged wavelet spectra of topographic
profiles extracted from four optical maps of stylolite surfaces.
Spectra have been normalized to superimpose for large $L$ on the
spectrum of S12A.}
\end{center}
\end{figure}

We analyzed the height distribution in terms of self-affinity
\cite{Barabasi95} which states that the surface remains statistically
unchanged for the transform: $\Delta x \rightarrow \lambda \; \Delta
x$, $\Delta y \rightarrow \lambda \; \Delta y$, $\Delta z \rightarrow
\lambda^\zeta \; \Delta z$, where $\lambda$ can take any real
value. The exponent $\zeta$ is the so-called Hurst exponent. A 1D
Average Wavelet Coefficient technique \cite{Simonsen98} has been
used. For a self-affine profile, the wavelet spectrum behaves as a
power law with a slope $1/2 + \zeta$, and provides an estimate of the
Hurst exponent $\zeta$. The spectra clearly exhibit two regimes
(Fig.~\ref{fig_spum_data}). At large length scales, a power law
behavior is observed with a slope of $1$ in the log-log plot, which is
consistent with a Hurst exponent of $\zeta_1=0.5$. At small length
scales, a second power law behavior is observed with a larger slope
($1.7$) in agreement with a Hurst exponent $\zeta_2=1.2$. The
crossover length scale is sharp and defines a characteristic length
scale which is slightly different for the three surfaces, $L_c\approx
1$ mm. $L_c$ is several orders of magnitude larger than the grain size
and significantly larger than experimental cutoffs. This spectral
behavior is observed for both mechanical and optical measurements.

\begin{figure}[ht]
\begin{center}
\includegraphics[width=12cm,angle=0]{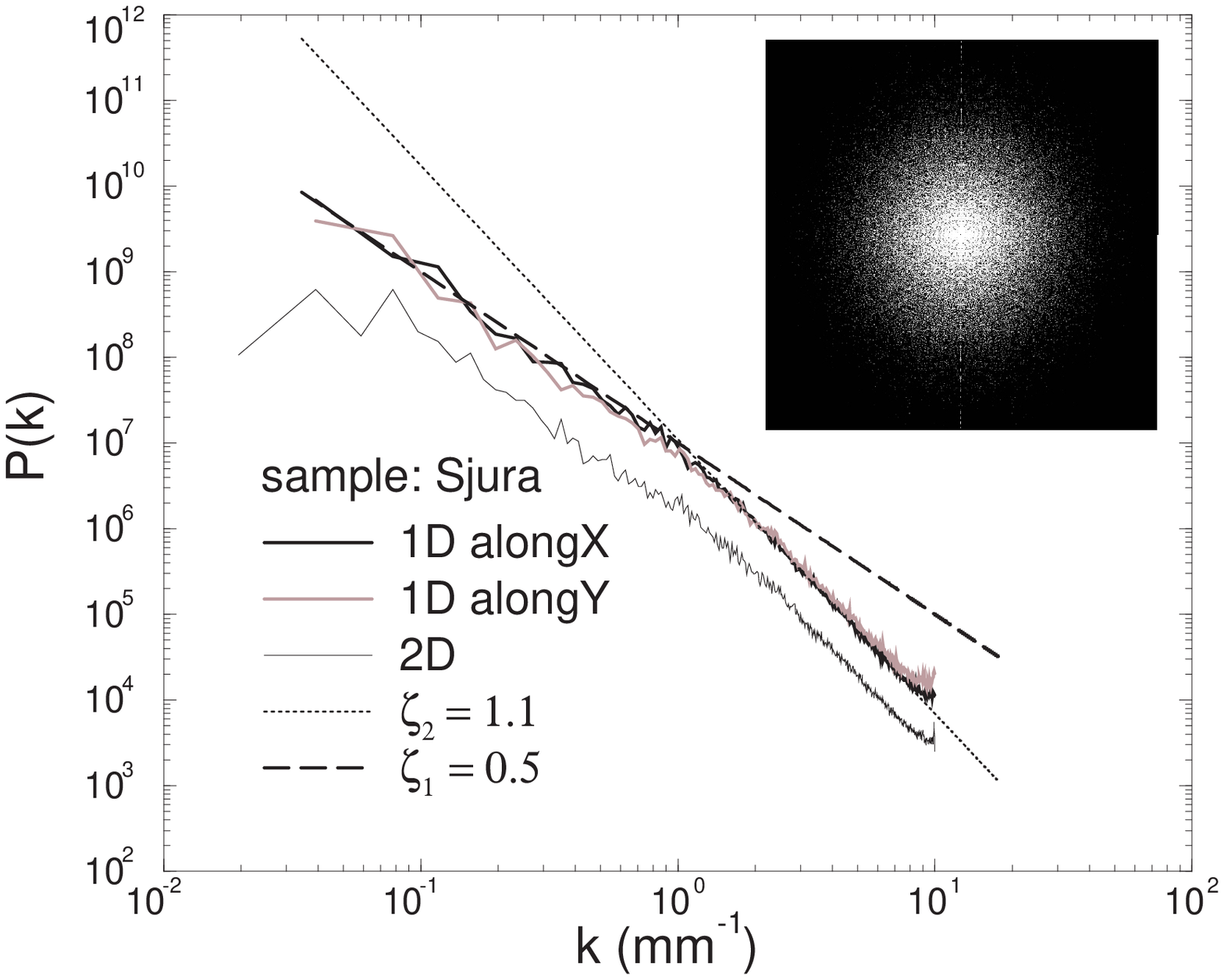}
\caption{\label{fig_spum_iso} Fourier power spectra of 1D topographic
profiles oriented along two perpendicular directions (X and Y) and of
the full 2D surface. The latter was radially integrated to be compared
to the 1D power spectra. Inset shows a gray map of the 2D power
spectrum. A mirroring technique has been used to reduce non periodic
edge effects.}
\end{center}
\end{figure}

We checked that another analysis technique, namely the Fourier power
spectrum, was providing very consistent
results. Fig.~\ref{fig_spum_iso} shows averaged 1D spectra of profiles
extracted from the surface Sjura. A self-affine property of the
profiles leads to a power-law behavior of the power spectrum as
$P(k)\propto k^{-1-2\zeta}$\cite{Barabasi95}. Moreover, average
spectra of profiles taken along perpendicular directions provide very
consistent results (Fig.~\ref{fig_spum_iso}). Isotropy of scaling
invariance is confirmed by the circular symmetry of the 2D power
spectrum of the surface.

The second part of the letter is devoted to a modeling of the
stylolite roughening. The aim is to understand the origin of the
self-affine behaviors and of the characteristic length $L_c$.  We
propose to consider a simple model as a paradigm for an interface
growth in an heterogeneous medium like natural rocks. This model aims
at providing a framework for future modeling of stylolite formation.


We consider the following geometry: The stylolite interface is assumed
to be initiated along the boundary between geological
beds. Accordingly it can be approximated as the boundary of a
quasi-flat and very elongated fluid pore. The trapped fluid is assumed
to form a film and to be at lithostatic pressure. The solid, where
this pore is embedded, is supposed to undergo an average stress
$\bm{\sigma^0}=\sigma^0_{zz} \hat{z}\hat{z} + \sigma^0_{xx}
(\hat{x}\hat{x}+\hat{y}\hat{y})$, where $\hat{z}$ refers to the
direction normal to average stylolite direction and $\hat{x}$ and
$\hat{y}$ refer to directions along the average stylolite
direction. Since stylolites are on average normal to the largest
principal stress direction,
$\sigma_s=|\sigma^0_{zz}|-|\sigma^0_{xx}|>0$.

Possible solid contacts with the mirror surface on the other side of
the fluid film are neglected, considering that such contact points
concentrate strain when they occur, and induce faster dissolution of
these contacts, thus leading to an essentially lubricated contact zone
between neighboring grains: for simplicity, we neglect interactions
between the mirror surfaces and assume that the front morphology is to
first order dominated by a dissolution process between a fluid film
and a single elastic solid.

Assuming a free surface profile $z(x,t)$, the normal $\hat{n}$ to the
interface pointing toward the solid is, in the limit of small relief,
$\hat{n}=\hat{z}-(\partial_x z) \hat{x}$, where we assume plane strain
perturbations.  The stress state in the solid is expressed as
$\bm{\sigma} = \bm{\sigma^0} +\bm{\sigma^1}$, where mechanical
equilibrium between solid and fluid requires that $\bm{\sigma^1}\cdot
(-\hat{n})= -\sigma_s (\partial_x z) \hat{x}$. This stress state
results from a surface distribution of tangential force -$\sigma_s
(\partial_x z) \hat{x}$ applied on the quasi-planar boundary of the
solid by the fluid, so that using Green's elastostatic function
\cite{Landau} and integrating along the $y-$direction, at the surface,
$\sigma^1_{xz}=\sigma^1_{zx}=\sigma_s (\partial_x z)$ and
$\sigma^1_{xx}=\sigma^1_{yy}=\sigma_s (2\nu/\pi)\int dy(\partial_{y}
z(y))/(x-y)$, all other components being null.

For small reliefs ( $||\sigma_1||/||\sigma_0|| \ll 1$) and to leading
order the elastic free energy $u_e=[(1+\nu)\sigma_{ij}\sigma_{ij}-\nu
\sigma_{kk}\sigma_{ll}]/4E$ can be approximated as $u_e=u_e^0+u_e^1$
where from the above,
\begin{eqnarray}
\label{u0} u_e^0&=&\alpha p_0^2/E\\
\label{u1} u_e^1&=&-\;\beta (p_0 \sigma_s/E) \int dy (\partial_{y} z)/(x-y)
\end{eqnarray}
with an average solid pressure
$p_0=-(2\sigma^0_{xx}+\sigma^0_{zz})/3$, and two dimensionless
positive constants $\alpha=[9(1-2\nu)+2(1+\nu)\sigma_s^2/p_0^2]/12$
and $\beta=\nu(1-2\nu)/\pi$, where $E$ is an effective Young's
modulus, and $\nu$ the Poisson coefficient. $u_e^0$ is the elastic energy
from the global tectonic loading and $u_e^1$ is its local perturbation that
results from the interface topography.

The chemical potential difference at the solid/fluid interface that
can potentially destabilize the interface can be written as
\cite{Kassner01}:
\begin{equation}
\label{eq_mu}
\Delta\mu = \Omega (u_e + \gamma\kappa)
\end{equation}
where $u_e$ is the elastic energy per unit volume in the solid,
$\gamma$ is the surface energy, $\kappa$ the curvature, and $\Omega$ a
molar volume. We have assumed that gravity effects are negligible.  We
have also assumed that the matrix of the solid, {\it i.e.} an assembly
of initial sedimentary particles, is sufficiently porous during
stylolites initiation to have a bulk diffusion within the material.
This assumption is supported by rock thin section observations under
an optical microscope \cite{Carrio92}.  If a bulk diffusion holds in
the fluid surrounding the stylolite, the evolution of the interface is
directly related to the chemical potential: $v_n=m\Delta\mu$ where
$v_n$ is the normal dissolution velocity and $m$ is the
mobility~\cite{Kassner01}. We also neglected the chemical potential
evolution within the film since we only aim at describing the
initiation of the process under drained conditions.

This homogeneous description thus predicts, for small reliefs,
$\partial_t z = v_0+m\Omega(u_e^1+\gamma \partial_{xx}z)$, with $v_0=m
\Omega u_e^0$. Surface tension is a stabilizing term, but it is
important to note that the elastic interaction term, $u^1_e$, is also
stabilizing in the present context.  For the present situation,
stylolites are perpendicular to the maximum principal stress, and will
subsequently be assumed horizontal: $\sigma_s>0$. Considering an
elementary departure from a flat interface, such as a fluid intrusion
in the solid, i.e. a bump with a maximum in x, such as $\partial_y z
>0$ for $y<x$, and $\partial_y z <0$ for $y>x$, $u_e^1$ is negative in
x and reduces the dissolution speed in the bump at x. Accordingly,
since the problem is linear, elastic interactions are stabilizing for
any corrugations of the interface. For vertical stylolites, the
picture is opposite ($\sigma_s<0$) and elastic interactions are
destabilizing leading to a lateral expansion of the stylolite.

The homogeneous picture predicts the propagation of a planar
dissolution interface driven by the average elastic energy $u_e^0$,
with an average speed estimated as $v_0 \approx 8\cdot10^{-6}$ m/year
where we used $m= k
\Omega/ (R T)$, with a dissolution rate $k\approx 10^{-4}$
mol/m$^{2}$/s, $\Omega\approx 4\cdot 10^{-5}$ m$^3$/mol for calcite,
$R$ is the universal gas constant, $T\approx 300$ K, $\alpha\approx
0.5$, $E\approx 8\cdot10^{10}$ Pa for limestones, a characteristic
stress estimated as $p_0\approx 25$ MPa, corresponding to a rock at 1
km depth.

To understand the dynamic roughening of stylolites, it is essential to
capture the effect of heterogeneities of relevant material properties
in the solid, namely $\nu,E,m$ and $\gamma$.  We assume the relative
variation ($\delta E/E$ and others) of these properties to be small,
and to correspond to independent random variables associated to each
constitutive grain of the rock, which are typically $\ell$=10 $\mu$m
sized. At early stages of the process where $\partial_x z\ll 1$, we
define the dimensionless surface position with respect to the average
plane $z'=(z-v_0 t)/\ell$ and the dimensionless space and time
variables $x'=x/\ell$ and $t'=t/\tau$ where $\tau=\ell^2/(\gamma
\Omega m)$ to obtain, to leading order in relative fluctuations and
typical slopes, for the roughening interface speed:
\begin{equation}
\label{eq:roughening_dyn}
\partial_{t'} z'(x',t')=  \partial_{x'x'} z' - \frac{\ell}{L^*} \! \int \!dy' \frac{\partial_{y'}z'}{x'-y'} + \eta(x',z'(x'))
\end{equation}
where $L^*=\gamma E/(\beta p_0 \sigma_s)$ and $\eta=[\alpha \ell
p_0/(\beta L^* \sigma_s)] \cdot [(\delta E/E) + (\delta m/m) - (\delta
\alpha /\alpha)]$.  In this Langevin equation with quenched noise, the
destabilizing random term is balanced by the restoring surface
tension term at scales below $L^*$, and by the restoring elastic
interactions at scales above $L^*$.  We propose that this critical
scale $L^*$ corresponds to the measured crossover length $L_c$.  For
typical limestones, $\gamma=0.27$ J/m$^2$ for a water-calcite surface
and $\nu\approx 0.25$, so that $\beta\approx 0.04$ and $L^*\approx 0.9
$ mm, consistently with the above measured. The other characteristic
quantities of interest are $\tau\approx 0.2$ year and the
characteristic amplitude of the dimensionless noise $\eta$ is $\rho
\approx \alpha \ell p_0/(\beta \lambda^* \sigma_s)\approx 0.2$.

For the Laplacian regime ($L \ll L^*$) and the mechanical regime ($L
\gg L^*$), only one of the two restoring terms in
Eq.~(\ref{eq:roughening_dyn}) dominates, and these two independent
regimes have already been studied. Indeed, the Laplacian regime is
nothing else than the Edwards Wilkinson (EW) problem \cite{Edwards82}
in a quenched noise. In this case the interface is self-affine with an
exponent $\zeta_2\approx 1.2$ \cite{Roux94}. In the mechanical regime,
Eq.~(\ref{eq:roughening_dyn}) is analogous to the quasi-static
propagation of an elastic line or a mode I fracture front in a
disordered material, and the Hurst exponent is $\zeta_1\approx 0.4$
for a kernel similar to Eq.~(\ref{u1})
\cite{Schmittbuhl95c,Ramanathan98,Tanguy98,Rosso02}.

The roughening amplitude can be obtained by considering the EW
equation with quenched noise regime: the characteristic width of the
surface measured at scale $L$ scales as $w(L)/\ell \approx \rho
(L/\ell)^{\zeta_2}$ at saturation, obtained from a flat interface
after a saturation time $\tau_s(L)$ such that $\tau_s/\tau\approx
(L/\ell)^{\zeta_2/\delta}$, with a dynamic exponent $\delta\approx
0.8$ \cite{Roux94}. With $L\approx 1$ mm, $\ell \approx 10 \mu$m and
$\zeta_2\approx 1.2$, this scaling law predicts up to a constant of
order unity the saturation width at cross over scale $w(L^*)\approx
0.5$ mm and the time to saturation as $\tau_s \approx 200$ years.
This length scale corresponds to the measured one
(Fig.~\ref{fig_roughness_prof}), and the short saturation time implies
that observed stylolites have achieved their saturation width over
geological time scales. That the width amplitude is also correctly
predicted in the mechanical regime could be checked directly, but is
granted by the fact that it is correctly predicted in the Laplacian
regime, as well as the crossover scale, which determines entirely the
prefactor of the scaling law $w(L)$ in the $L>L^*$ regime.  In
principle, determining $L^*$ and $w(L^*)$ could give two independent
constraints on both $p_0$ and $\sigma_s$, which could allow to
determine both the pressure and differential stress prevailing during
the formation of a particular stylolite.  However, given the
amount of approximations in the involved constants, the only way to
test this effect on the cross-over wavelength would be to measure
stylolites formed in various geological conditions and study the
effect of depth and orientation to the main stress.


In conclusion, we presented a quantitative description of stylolite
interfaces. The experimental measurements are 3D high resolution
descriptions of the topography of natural stylolites. We show that the
surfaces are self-affine but with two regimes. At small scales, the
Hurst exponent is unexpectedly high, $\zeta_2=1.2$, and consistent
with a Laplacian regime. At large scales, the stylolites morphology is
controlled by long range elastic stress redistributions. In this case
the roughening is important with a low Hurst exponent $\zeta_1=0.5$.
The two regimes are separated by a crossover characteristic length
$L_c$, also predicted by a model based on the description of a
stress-induced dissolution, where restoring surface tension effects
and elastic interactions compete with a quenched noise.  It is
important for geological implications to note that $L_c$ is very
sensitive to the average stress $p_0$. Indeed, a measurement of $L_c$
from roughness profiling could provide an estimate of the stress
magnitude during the stylolite growth, that is, in the
past. Accordingly stylolites could be considered as stress fossils.

We acknowledge D. Rothman, J. Rice, A. Lobkovsky, B. Evans,
Y. Bernab{\'e}, B. Goff{\'e}, H. Perfettini, P. Meakin, and E. Merino
for fruitful discussions and two anonymous reviewers for their 
constructive comments.

\bibliographystyle{prsty}

\end{document}